# "Welcome to Gab": Alt-Right Discourses

Nga Than, City University of New York – The Graduate Center, nthan@gradcenter.cuny.edu
Maria Y. Rodriguez, University at Buffalo, myr2@buffalo.edu
Diane Yoong, City University of New York – The Graduate Center, dyoong@gradcenter.cuny.edu
Friederike Windel, City University of New York – The Graduate Center, fwindel@gradcenter.cuny.edu

## ABSTRACT

Social media has become an important venue for diverse groups to share information, discuss political issues, and organize social movements. Recent scholarship has shown that the social media ecosystem can affect political thinking and expression (Bail et al. 2018; Berger 2018; Lewis 2018). Individuals and groups across the political spectrum have engaged in the use of these platforms extensively, even creating their own forums with varying approaches to content moderation in pursuit of freer standards of speech. The Gab social media platform arose in this context (Lima et al. 2018). Gab is a social media platform for the so-called "alt-right," and much of the popular press has opined about the thematic content of discourses on Gab and platforms like it, but little research has examined the content itself. Using a publicly available dataset of all Gab posts from August 2016 until July 2019 (Baumgartner 2019), the current paper explores a 5% random sample of this dataset to explore thematic content on the platform. We run multiple structural topic models (Roberts et al. 2014), using standard procedures to arrive at an optimal k number of topics. The final model specifies 85 topics for 403,469 documents. We include as prevalence variables whether the source account has been flagged as a bot and the number of followers for the source account. Results suggest the most nodal topics in the dataset pertain to the authenticity of the Holocaust (topic 19), anti-feminist political philosophy (the meaning of red pill, topic 27), and the journalistic merit of the so-called "mainstream media" (topic 35). We conclude by discussing the implications of our findings for work in ethical content moderation, online community development, political polarization, and avenues for future research.

## KEYWORDS
Social media, alt-right, content moderation, structural topic modeling, political polarization

## INTRODUCTION

The ecosystem of social media platforms such as Twitter, Facebook, and YouTube has increased the ease of global communication, particularly opinion sharing. As a movement that predominantly emerged online (Berger 2018; Heikkilä 2017), the alt-right,[1] an umbrella term for loosely affiliated movements across the world mainly centered in the United States (Berger 2018), has been proficient at utilizing and leveraging digital tools for recruitment and propaganda (Daniels 2018). Its propensity for adopting these technologies to reach out and share its ideologies with different publics and communities (Lewis 2018; McIlroy-Young and Anderson 2019) is connected to a much longer legacy of white supremacy (Daniels 2018). Being early adopters of emerging technology, the alt-right is primed to interact with younger users who often spend significant time on these platforms (Lewis 2018). Recent features in the *New York Times* (Roose 2019) and *Washingtonian* (2019) have shown that younger white men on various social media platforms have been "radicalized" by the rhetoric of the alt right. In part as a response to the alt-right's social media presence and reach, several mainstream social media platforms have created stronger content moderation policies. The result has been the concurrent abandonment of mainstream social media platforms and the development of more content-permissive platforms, like Gab (Berger 2018).[2]

[1] The term alt-right is sometimes used interchangeably with "white supremacy," "neo-Nazi," and "fascist." While they share historical legacies and genealogy there is some distinction between these groups (Berger 2018, Stern 2019). We choose to use the term "alt-right" to represent the discourses on Gab, believing them to be distinct from mainstream white supremacy, similar to what has been done.

Gab describes itself as "a social network that champions free speech, individual liberty, and the free flow of information online. All are welcome" (https://gab.com). Gab emerged in response to the banning of several prominent personalities on Twitter, following a series of public events in which users were promoting hate-speech (Berger 2018). Gab's promise of free speech translates into little content moderation (Lima et al. 2018) and has attracted many users whom other social media platforms have banned. As such, many alt-right supporters have chosen to become users of this social media platform, resulting in an "echo-chamber" in which users are predominantly fed with information that reinforces their own opinions and ideological views (Lima et al. 2018;

Thomas 2018). Gab users are fed with and supported by information that serves their own alt-right views. While research on alt-right conversations in mainstream social media platforms like Twitter (Berger 2018) and YouTube (Lewis 2018) exists, less is known about the discourses circulated on Gab, which this paper investigates. We employ structural topic modeling (STM), an unsupervised machine learning technique for text classification, to identify broad topics across a random 5% sample of Gab posts between August 2016 and July 2019. This paper analyzes the qualitative distribution of themes within Gab posts to develop potential content moderation strategies ideal for deterring political extremism.

## RELATED WORK

### *Social Media and Political Polarization*

Debates on the potential of social media interaction to unify or divide have been part of research and everyday conversations since the beginnings of the Internet. In 2014, the Pew Research Center found that ideological uniformity among Americans with both conservative and liberal opinions doubled within two decades (Marwick and Lewis 2017). This finding corresponds to the increasing use of social media as well (Berger 2019). Recent scholarship indicates that Gab has become more polarized since its inception in 2016, with an increase in alt-right topics over time (McIlroy-Young and Anderson 2019). The findings suggest Gab users initially engaged in community-building discourses, then shifted to alt-right political topics including increases in anti-Semitic language. In response to concerns about rising political polarization due to "echo chambers" that "create information bubbles" (Lima et al. 2018) and limit exposure to opposing views, Bail et al.'s study examined the effects of exposing twitter users who identified as Democrats or Republicans to opposing views (2018). The authors found the exposure is in fact counterproductive; it increased political polarization, specifically for Republican followers, who grew more conservative in response to opposing views. While the focus of this study was party-identified Americans and thus cannot be generalized, the authors refer to recent studies that indicate the influence the politically conservative population has on public debates. Our research raises important questions about how to reduce political polarization in online spaces, possibly by learning about the kinds of messages or framing of positions that backfire and those that can address political polarization.

### *Social Media and The Alt-right*

When Paul Gottfried, a humanities professor, coined the term alternative right in a speech in front of other right-leaning intellectuals in 2008 (Marwick and Lewis 2017; Hartzell 2018; Stern 2019), the ecosystem of social media platforms was just emerging. Following this speech, Gottfried formed an "independent intellectual right" organization with Richard Spencer, a self-identified white nationalist, who two years later would launch AlternativeRight.com (Hartzell 2018), which aimed to address the concerns of the "white nationalist intelligentsia" and develop a plan to grow the movement by reaching a younger audience. Spencer took on this role and moved online to reach a wider audience for the alt right (Hartzell 2018) whose activity on social media platforms like Twitter and Facebook has brought its discourse into public conversation (Heikkilä 2017; Stern 2019).

Daniels stresses the alt-right as both an organizing force that manifests white nationalism and white supremacy and an influence on internet culture, described as "the emerging media ecosystem powered by algorithms" (Daniels 2018). While white nationalists strategically spread anti-Semitic and racist ideologies through social media, algorithms have helped amplify their spread. She argues the need to investigate the relationship between hate crimes, white nationalism, and the algorithms of search engines and social media platforms. Marwick and Lewis (2017) have investigated the strategies that the alt-right employs in this current media system, including the strategic use of memes and bots, as well as using their knowledge of internet culture and irony. Like Daniels, they look at both the organizations behind the spread of racist and anti-Semitic ideologies and the current media ecosystem. Vidgen's work on Islamophobia on Twitter demonstrates the way in which this platform, with its limited character count, encourages "aggressive antagonistic interactions" rather than democratic dialogue (2019). In a late 2018 report, Lewis describes the manner in which YouTube rewarded content producers and their associated networks regardless of their content, thereby rewarding alt-right influencers using this business model, which not only generated a community of followers for these influencers but also provided them with monetary incentives to generate larger followings. In their solution-oriented paper on sociotechnical security, Goerzen et al. (2019) frame social media platforms as sociotechnical systems that bring human and non-human actors together and therefore are governed not just by social but also technological norms.

### *Shared Thematic Concerns Across the Alt-right*

The alt-right movement is created through the spread of white nationalist and white supremacist ideologies and strategic use of the social media ecosystem. Similar to Daniels's concerns about framing the alt-right merely as the angry white male, and thus around individual actors with extremist perspectives, this paper approaches the alt-right as creators of a network of communities that utilizes various strategies and tools to produce and maintain discourses and activities (Daniels 2018; Marwick and Lewis 2017). Looking at alt-right Twitter users, Berger has identified four discourses that circulate among alt-right users including support for Trump and white nationalism as well as anti-Muslim speech (Berger 2018). The main influencers that Berger found in this network often identified as white nationalists, demonstrating the white nationalist sentiment that underlies alt-right movements (Daniels 2018). Lewis points to the ways in which, on YouTube, content creators on the right have created an alternative influence network (AIN) that is a "fully functioning media system" offering alternative news and media to the mainstream (Lewis 2018). The AIN takes advantage of YouTube's Partner Program (YPP), where they are financially rewarded for their ability to broadcast and build communities. Through this process, AIN facilitates radicalization on the part of content creators and their viewers. On mostly alt-right social media platforms such as 8chan and Gab, they have hosted white terrorists who were spreading their manifestos prior to their shootings (Thomas 2019, McIlroy-Young). These shooters were cultural producers of alt-right content whose messages were retweeted, reposted and circulated. Understanding the alt-right as a community in which actors strategically use media to spread their ideological discourses requires the examination of the communicative practices of the alt-right (Nadler 2019).

**Production and maintenance of culture**

While Lima et al. (2018) note that the 2016 U.S. presidential election motivated a large group of new users to join Gab, the platform has since attracted a fair number of users outside of the United States. As such, the rise of the global right has created relationships with U.S. alt-right movements, and while race-based nationalism is still highly relevant in unpacking the ideologies driving these discourses, a broader set of frameworks is needed to understand the relationship between these communities. Davis points to the ways in which these communities can transcend geographic limitations and gestures towards a transitional movement that is paradoxically in "defen[s]e of national and other chauvinisms" (Davis 2019). Using Australian extremist groups as an example, he articulates how they are nationally based but share transnational ideologies with common focuses on anti-refugee ideologies and anti-Semitism. In his study of Facebook posts, he notes the wide circulation of images and memes that also include a modified version of Trump's campaign slogan and the use of Pepe the frog.

Davis notes the ways in which publicity, authority, and accountability to sovereign government changes as the movement becomes transnational (Davis 2019). Similarly, Heikkilä (2017), and Stern (2019), note that alt-right movements are largely decentralized and ideologically diffused. Through this fluidity and movement, the production of cultural objects becomes a way to engage with and sustain the alt-right. Both scholars point to the 2016 U.S. presidential campaign influential in the ways that alt-right cultural objects, such as memes and troll culture, engage with mainstream discourse. Both trace these "new" objects to the longer history of white nationalism. Heikkilä (2017) points to the ways in which white nationalism has always produced its own cultural objects such as films, music, and video games in opposition to the mainstream, while Stern uncovers the way in which the alt-right remixes and appropriates mainstream cultural products to facilitate its own retellings (Stern 2019). Broadly, the alt-right engages with metapolitics, that is, it is more interested in the production of social norms and cultural products rather than a direct engagement with politics. Hence, Clinton's campaign strategy that brought out the "deplorables" provided the alt-right the necessary attention and engagement with the hegemonic public that fueled, and continues to fuel, its strategies and movements.

Heikkilä (2017), Stern (2019), and Gökarıksel et al. (2019) reflect an urgency in studying the right. The tension between giving these movements attention or not has real risks and repercussions that are often violent and often happen in real life. Understanding the topics that proliferate in these movements is a first step in interrupting recruitment and disassembling the movement. Looking into the Gab echo-chamber provides us with foundational knowledge with which to begin this work.

## DATA & METHODS

Our main dataset is derived from Gab, an online social media platform whose functionality is a hybrid of Facebook and Twitter (McIlroy-Young and Anderson 2019). Recent scholarship has shown that Gab attracts far-right users because of its minimum content moderation policy (Lima et al. 2018). Users can post content, follow other users, and repost a status. They can also organize a post's topics with hashtags, a function that both Facebook and Twitter have.

We are interested in the diversity, or qualitative distribution, of topics that users have posted from the inception of the platform. Our dataset is comprised of all Gab posts from October 2016 to July 2019. In total we collected over 10 million posts. The data is publicly available and can be downloaded from pushshift.io (Baumgartner 2019). This paper utilizes a five percent random sample of the entire dataset. We then focused only on posts in English by comparing the language variable provided by the dataset, as well as running a cdl2 algorithm, a Google dictionary language detection algorithm. We keep only posts that are in English detected by both the Gab platform language function and the Google dictionary algorithm. Following McIlroy-Young and Anderson (2019) we removed spam posts and the automated "Welcome to Gab.ai" message which all users receive once they join the platform. Our final data are comprised of 403,469 posts posted by 40,375 users.

### Structural Topic Modeling

Previous research looking at discourses on Gab uses trigram and toxicity scores (Lima et al. 2018; McIlroy-Young and Anderson 2019). We employ structural topic modeling (STM) (Roberts et al. 2014) to investigate the different topics that users discuss on Gab platform. One advantage of STM is that it allows for contextualizing a topic in different contexts. One can examine the content of posts within a topic and the content of posts whose topics are correlated with others. This multi-level contextualization gives insights on text data that other automated text analysis methods cannot offer. Our work is the first to use STM to analyze the Gab dataset, and our results conclusively categorize Gab as an alt-right platform. In contrast, prior methods used simple term frequency or toxicity metrics. Our use of STM is motivated by its success in analyzing many political and social datasets (Nelson et al. 2018; Rodriguez & Storer 2019).

We used structural topic modeling (STM) by Roberts et al. (2014) to understand discourses produced on Gab. First, we conducted pre-processing standardization of the text by treating terms within documents as "bags of words," where each term represents a single feature and information on word order is discarded. Terms were also reduced to their stem form, such that "immigration" and "immigrating" become a common feature "immigrat." Additionally, stop words with no sematic meaning such as "the" and "of" were removed from the corpus. Then, we applied structural topic modeling by Roberts et al. (2014) to determine the topics within the corpus. Topic models are unsupervised machine learning models that automatically discover latent topics from text documents. In these models, a topic can be understood as a set of words representing interpretable themes, and documents are represented as a mixture of these topics. For each document, proportions across all topics sum up to 100%. For example, after fitting a topic model, a document can mostly be captured in the topic "American first" (proportion 70%), followed by "border control" (20%), and other topics (10%). In addition to representing documents as a distribution of topics, structural topic models further allow the inclusion of meaningful document-specific covariates that affect both document-topic proportion and word distribution. Drawing on this feature, we incorporated follower counts, and whether a post was flagged as a bot-written status as explanatory covariates to analyze how topic proportions vary.

While topic models are useful for reducing the dimensionality of textual data, one challenge is that the analyst must choose the number of topics in advance. We performed a searchK function provided in the stm package in R (Grimmer & Stewart 2013), and the results indicate that a model with 85 topics would be most appropriate because it would ensure the highest semantic coherence, lowest residual, and a high upper bound (See Figure A1 in Appendix A). We also ran multiple models from 15, 20, 25, 30, 35, 40 to compare the different aspects of the models (Figure A2 in Appendix A). It showed that a model with 85 topics indeed ensures more semantic coherence than other models. Finally, we used the package stm_insights to visualize the topic correlations and examine each topic in depth (Schwemmer 2018). The results will be presented in the following section.

One concern when employing topic modeling is how to calculate the number of topics to be defined by the model (Grimmer and Stewart 2013). Researchers decide based on some empiric motivation. While there are methods for estimating the likelihood that the number of topics selected is correct, there is currently no scientific consensus on how best to select the number of topics a model defines (Grimmer and Stewart 2013). Advances in computational social science, however, have made the selection process much more data-driven (Roberts et al. 2014). The STM package uses an algorithm developed by Lee and Mimno (2014) to estimate the number of topics based on a spectral initialization of the model. Though this method provides no statistical guarantees, it is a useful starting point for topic selection.

This article begins analysis using Lee and Mimno's (2014) topic selection algorithm on the entire random sample and offers goodness of fit diagnostics to justify the final number of topics selected. Further, we use the prevalence variable specification features of the STM model to allow topics to vary based on the number of followers a Gab user has and whether the network has flagged the Gab account as a bot.

**RESULTS**

The current study fits four STM models to the data (Roberts et al. 2014). The preprocessing functions of the STM package removes stop words, punctuation, and capitalizations for all posts (Roberts et al. 2014). Diagnostics for selecting the number of k topics are given in Figure 1, below. Held-out likelihood refers to the log probability of topics in the test set correctly replicating topics in the training set. The lower bound refers to the lower bound of the marginal log likelihood. Residuals refers to the difference between expected and predicted topic predictions. Semantic coherence refers to the co-occurrence of words in a topic – where co-occurring words also comprise the most likely words of a given topic, that topic has high semantic coherence. Figure 1 shows that between 80 and 90 topics produce relatively low residuals and a maximized lower bound, though held-out likelihood and semantic coherence is not optimal. Eighty-five topics were selected as the final K for the current data set and model.

Concurrent with k selection diagnostics, several test models are also run (Figure 2). Model 2 uses Lee and Mimno's algorithm (2014) to select k, with no prevalence variables indicated. Model 3 uses the k selection algorithm again but specifies whether the account is a bot and the number of followers as prevalence variables. The final model, model 4 specifies 85 topics and both aforementioned prevalence variables. Figure 2 indicates model 4, of all the models presented, offers the best model fit for the current dataset.

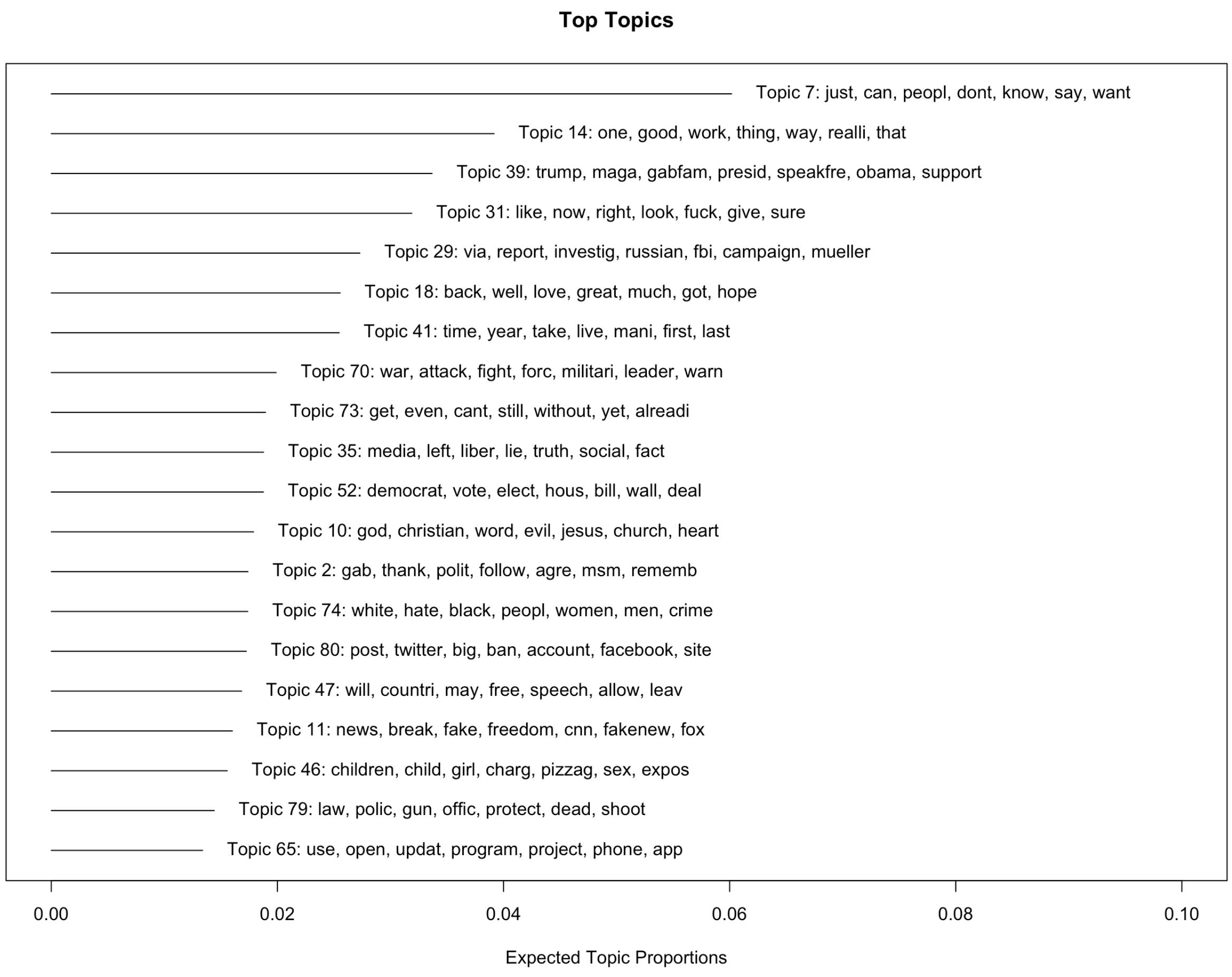

Figure 3: Top 21 topics by prevalence

Figure 3 offers the top 20 topics estimated by model 4. The figure offers the seven most frequent words defining a topic within a model, and the proportion of the dataset that contains the topic.

The first three topics make up approximately 14% of the dataset: topic 7, topic 14, and topic 39. Figures 4-6 offer visualizations of the top 3 topics, showing two example Gab posts and a word cloud of the most frequent terms within the topic.

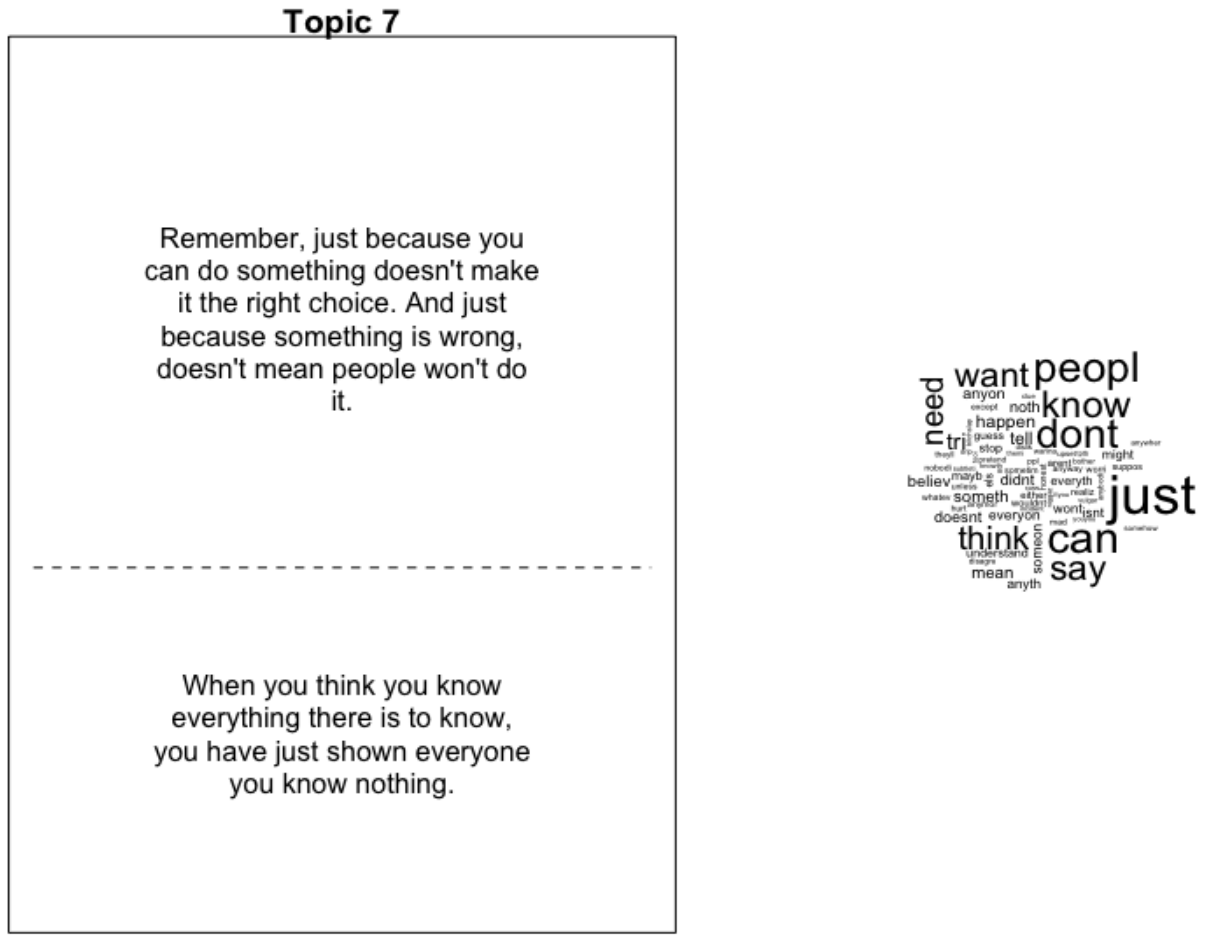

Figure 4 Visualization of Top Topic: Topic 7

Topic 7, illustrated in Figure 4, is suggestive of a “tendency toward intellectualism“ of the alt-right (Stern 2019). The seven most frequently occurring terms defining the topic – *just, can, people, don’t, know, say, want* – suggest the main content of the posts is users expressing their knowledge, opinions, or beliefs on the platform. This corresponds to studies that noted the academic origin of the term “alternative right” (Marwick and Lewis 2017; Hartzell 2018; Stern 2019). The content also suggests a tension between morality and wisdom in so far as morality is a gray affair and wisdom involves a level of humility that some eschew. Recent research (Schradie, 2019) finds that “freedom of speech” is used as a rallying point for right-leaning groups to mobilize supporters. These groups receive substantial foundation financial support to train personnel to produce content about freedom of speech online. In light of recent scholarship, we do not assume that topic 7 is a benign discussion of freedom of speech but rather a coded political message for mobilizing users to fight for their freedom of speech, which is allegedly being taken away by legacy media and big tech companies through content moderation.

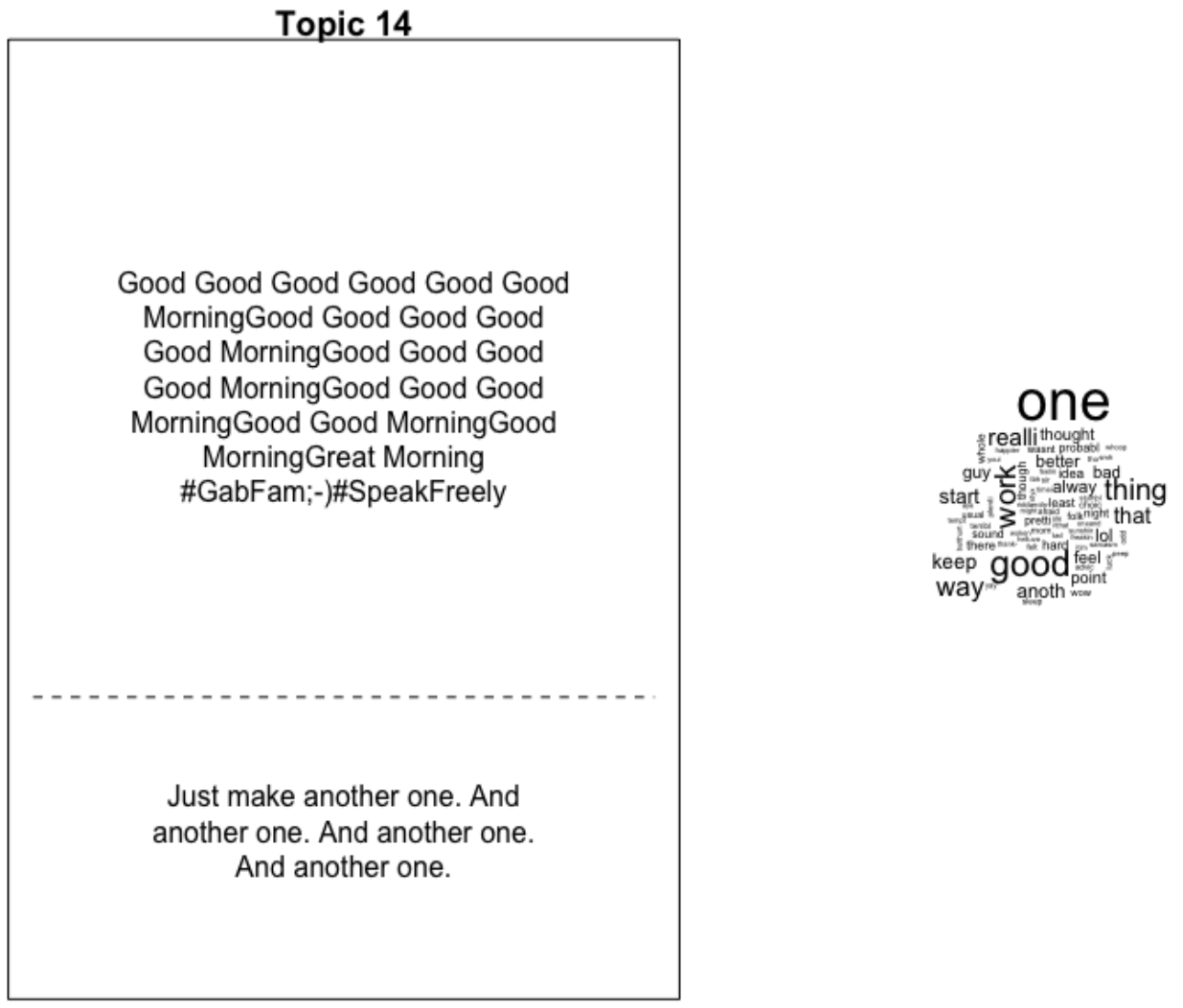

Figure 5: Visualization of Top Topic: Topic 14

Figure 5 visualizes Topic 14. The seven most defining words of the topic are *one, good, work, thing, way, really,* and *that.* The topic centers on the issue of freedom of speech that Gab promises its users. The ability to speak freely within Gab is important to users, especially when other social media platforms have chosen to ban or restrict user content (Berger 2018, 2019). Users also create a virtual community through the hashtag #Gabfam.

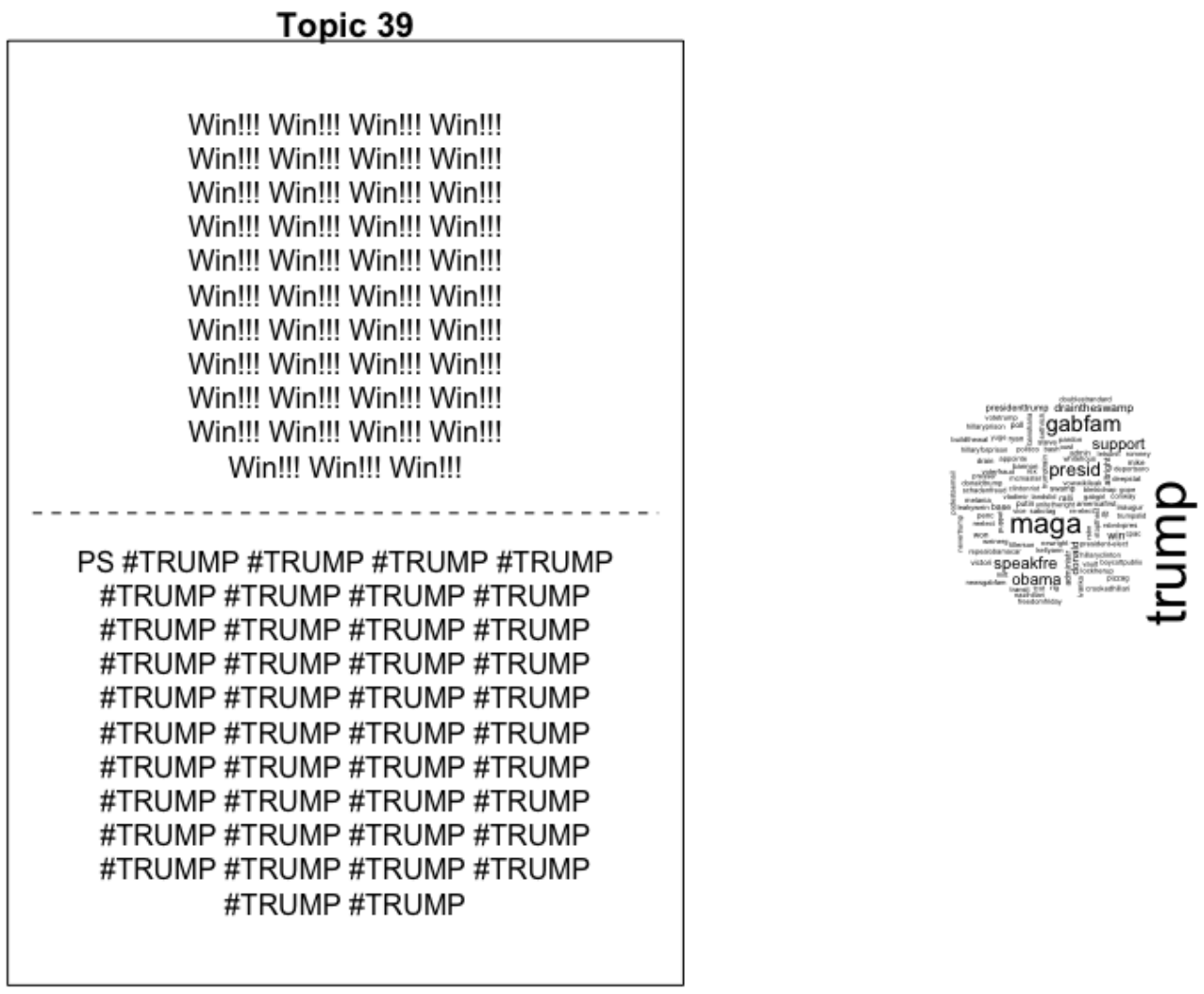

Figure 6: Visualization of Top Topic: Topic 39

Topic 39 shown in Figure 6 is defined by the seven most frequently occurring terms: *trump, maga, gabfam, president, speak free, obama,* and *support.* The topic appears to be centrally related to political discourse. Similar to what (McIlroy-Young and Anderson 2019) found, this topic shows the prevalence of political discussions on Gab, especially in support of Trump. Trump's election was a transitional moment for the alt-right (Heikkilä 2017), which is similarly reflected on the platform.

To summarize, the three top topics estimated by the final k = 85 topic model suggest that the majority of posts in the sample draw Gab users' attention to the platform's freedom of speech/no-to-minimal content moderation policy, and political discussions about the Trump presidency. To further investigate the qualitative distribution of topics, we employ STM Insight R package (Schwemmer 2018) to examine the correlations of topics as well as the estimated effect of prevalence variables.

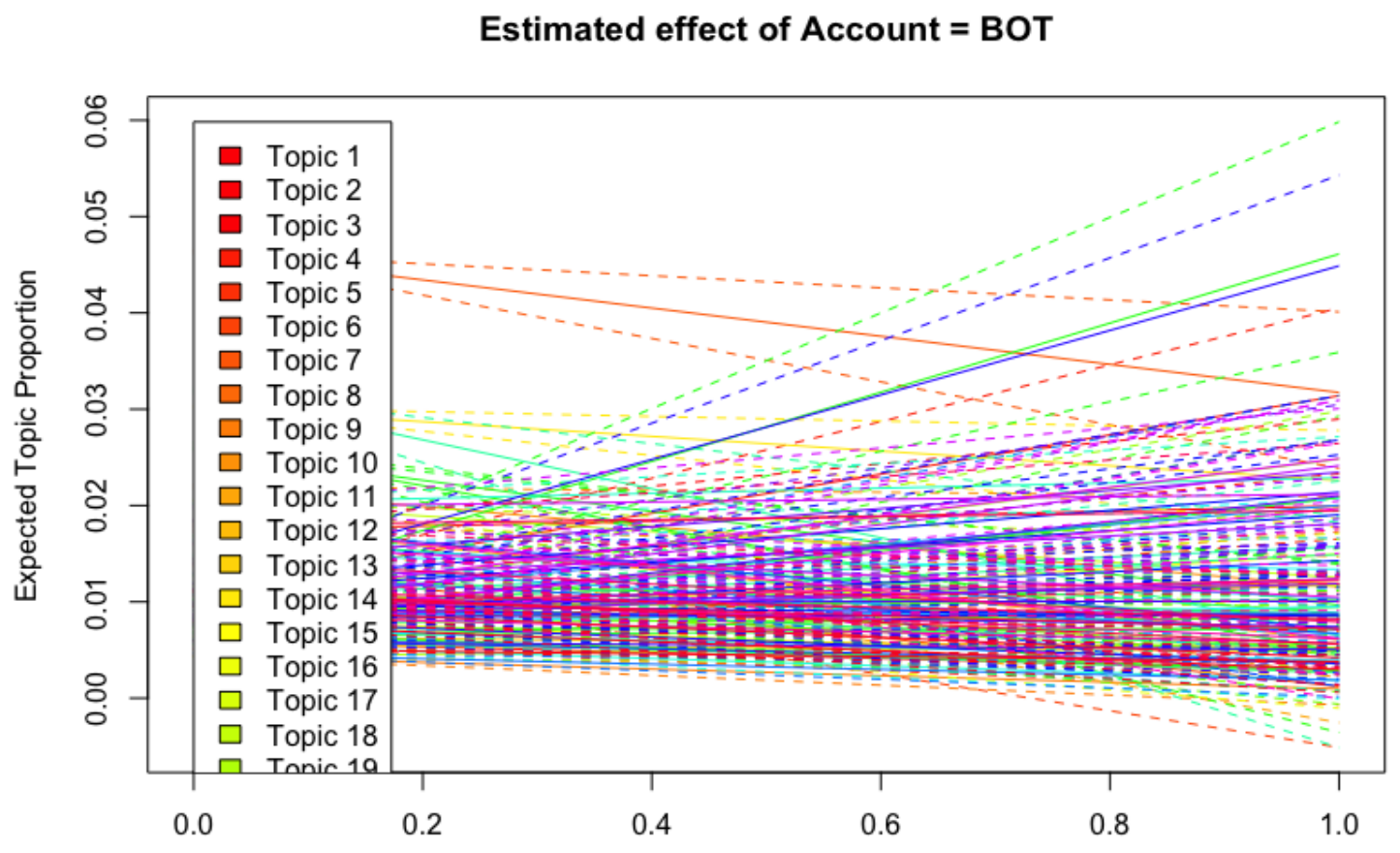

Figure 7: Estimated effect of Account flagged as bot on topic prevalence

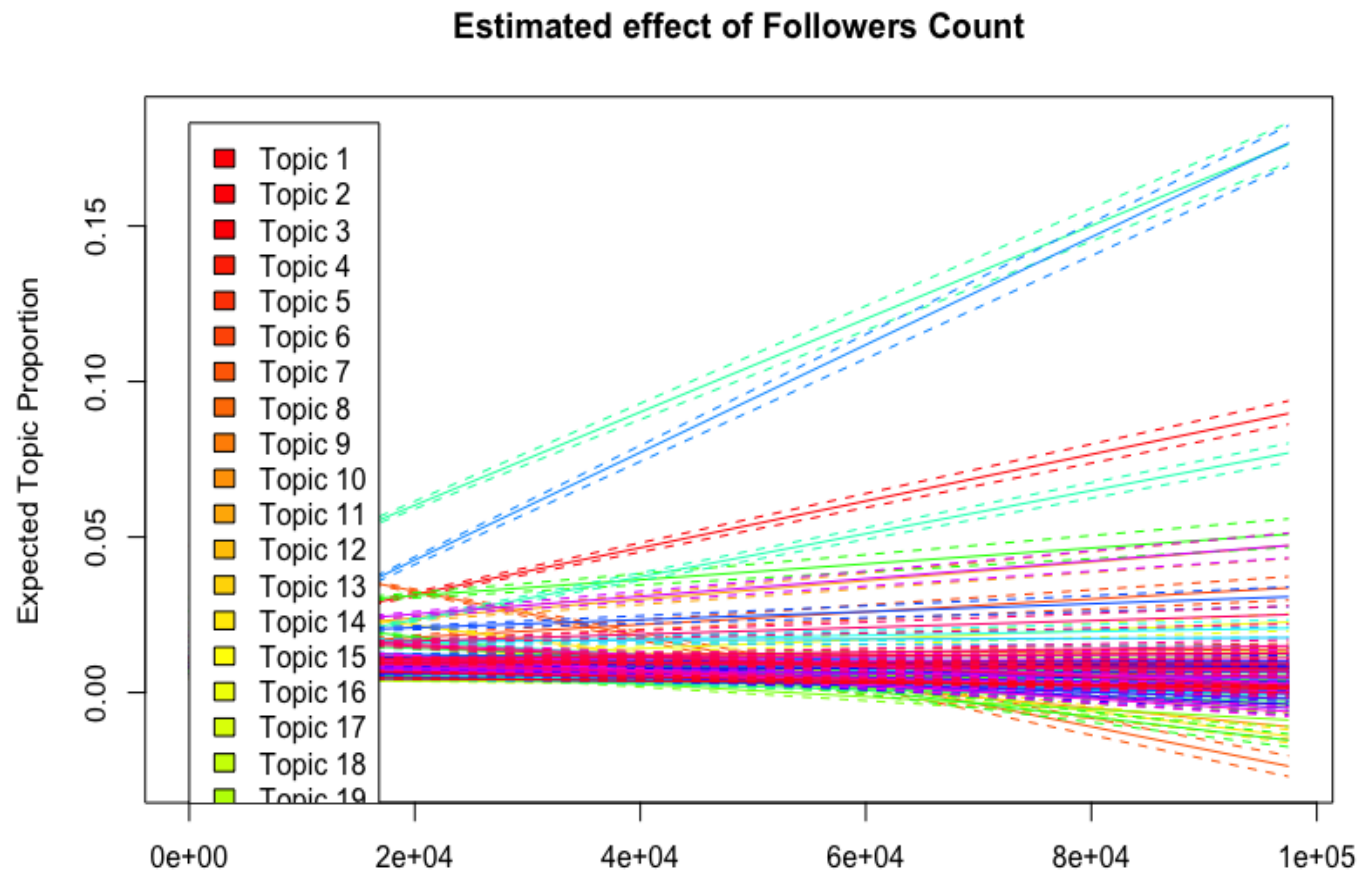

Figure 8: Estimated effect of follower count on topic prevalence

Figures 7 and 8 above indicates the estimated effect of an account being flagged as a bot (Figure 7) and the number of followers an account has (figure 8) on the topic distribution. Topics 12 and 13 appear to be most centrally related to bot accounts, while Topics 11, 13 and 14 appear to be most related to accounts with large number of followers. Topic 13, flagged in both graphs, is visualized below:

Bots are a new research area, and how they can influence people's perception about truth and information online. Our result shows that bots play a role in in information sharing, and community forming on Gab.com.

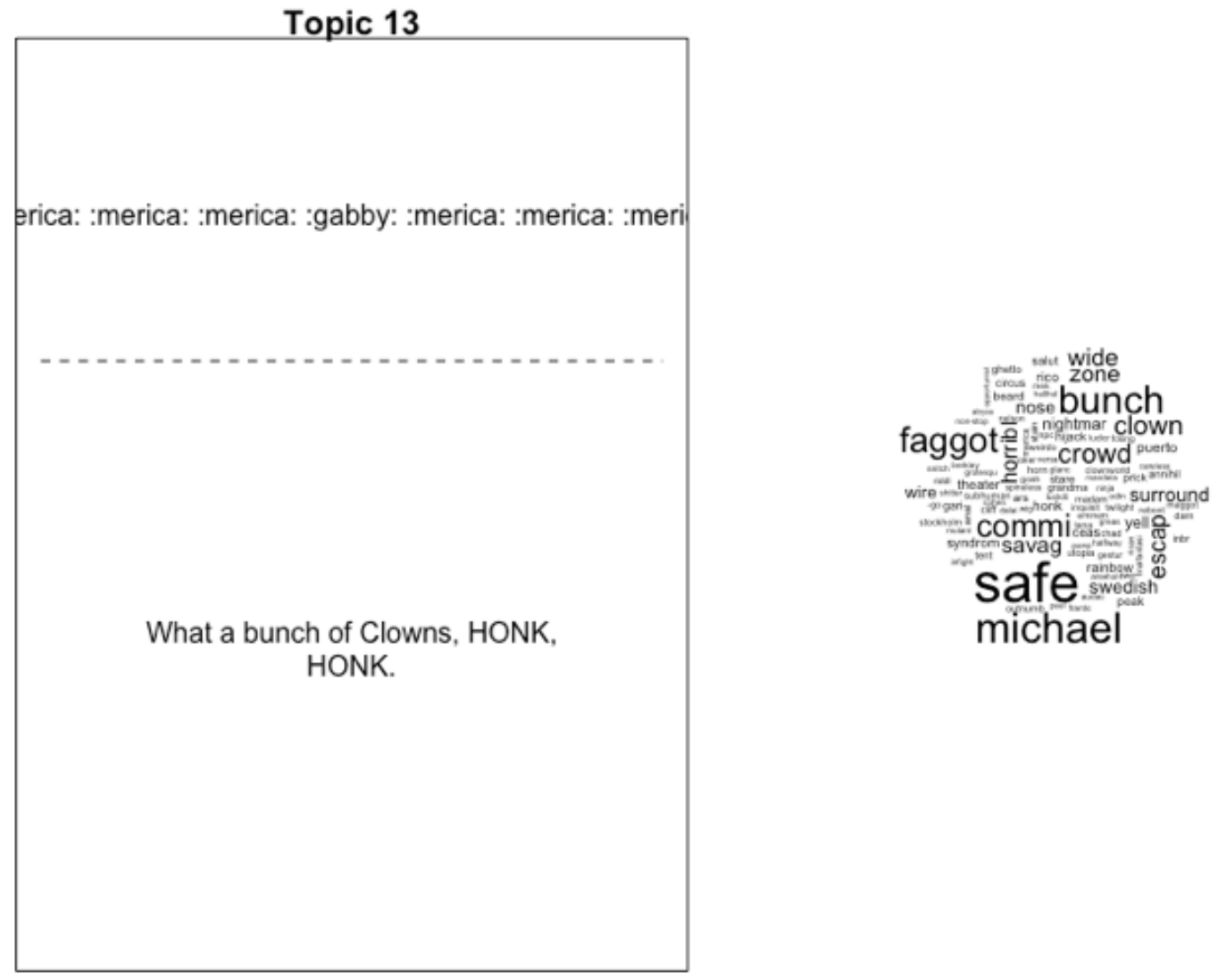

Figure 9: Topic 13

The topic lacks real coherence, indicating that it is perhaps the result of bot reposting behavior, or a natural language processing algorithm that takes a certain percentage of trending topics and produces content on them.

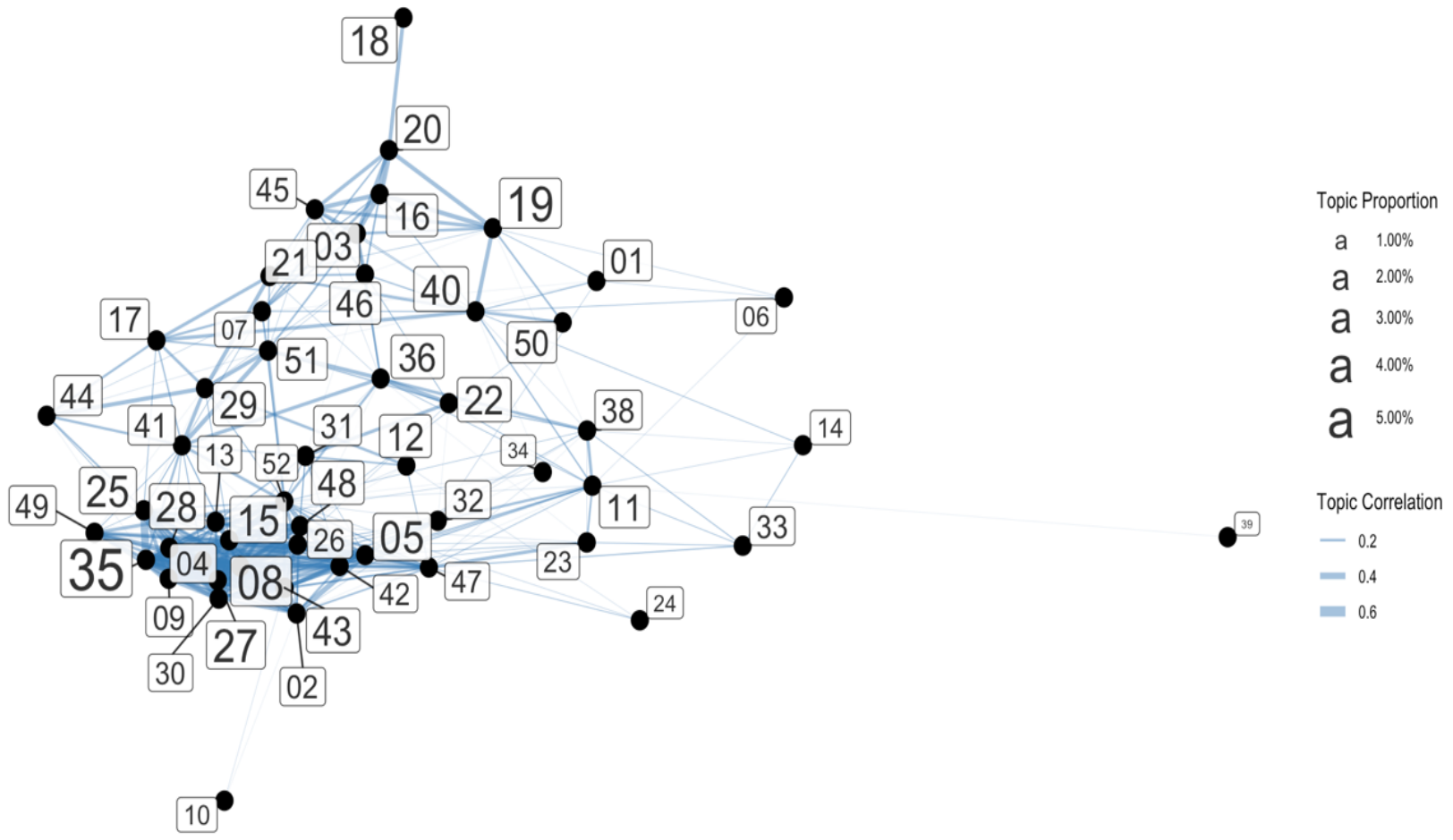

Figure 10: Topic Correlation Network

Figure 10 illustrates a network analysis of topic correlations from the final topic model produced with the STM Insights package (Schwemmer 2018). The larger the topic number, the higher the proportion of the data is comprised of the topic. Thicker lines between topics suggest stronger correlations between them. The figure indicates topics 19, 27 and 35 are nodal to the entire sample. Figures 10, 11 and 12 visualize each in turn.

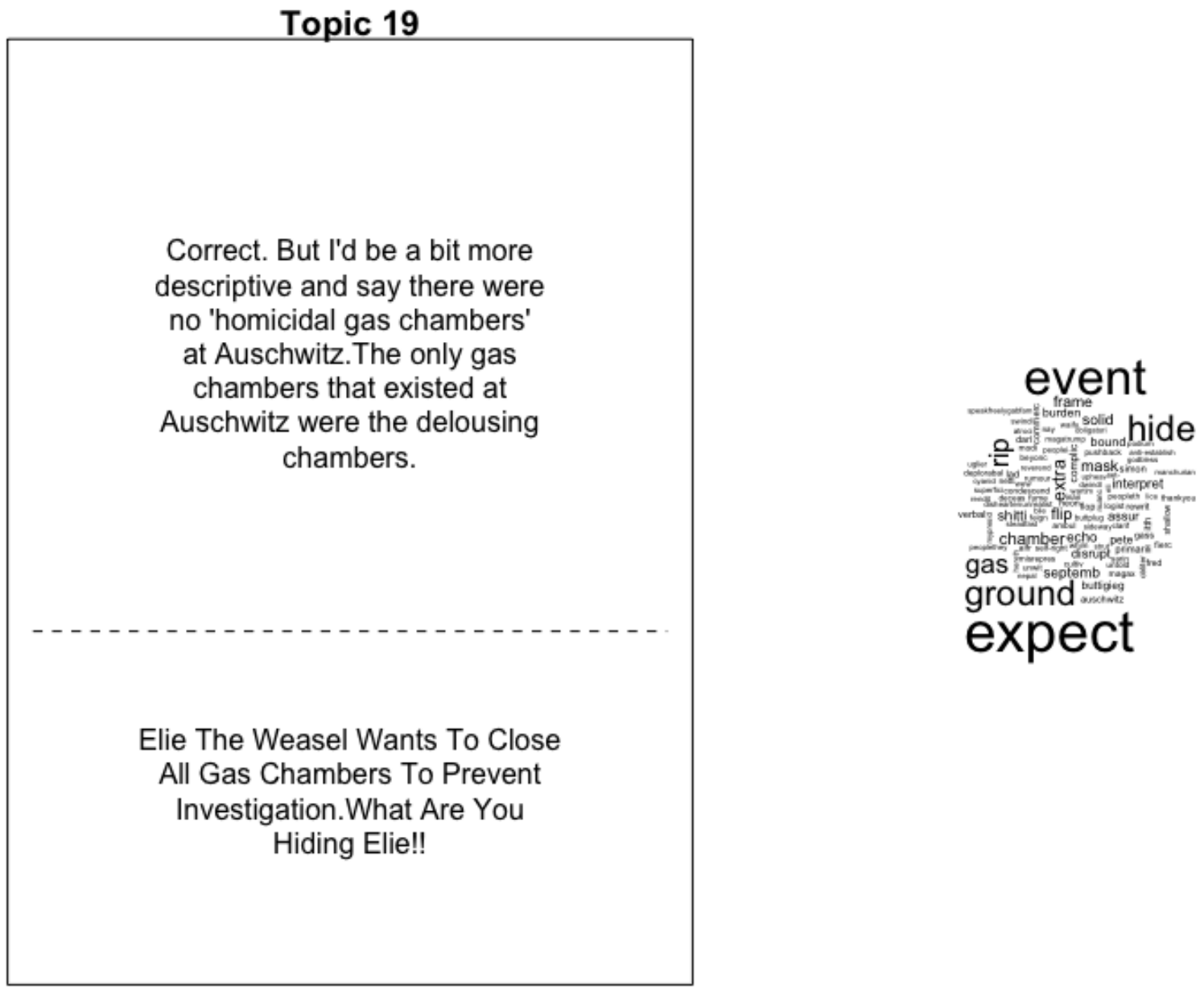

Figure 11: Visualization of nodal Topic 19

Figure 11 visualizes topic 19, which is a theme concerning the so-called fabrication of the Holocaust. This topic is a key area of concern for content moderation, content that is verifiably un-true yet propagated by network users. Silencing the content related to this topic plays into mainstream media tropes of political extremism but allowing it to flow unimpeded through a network is at best undesirable.

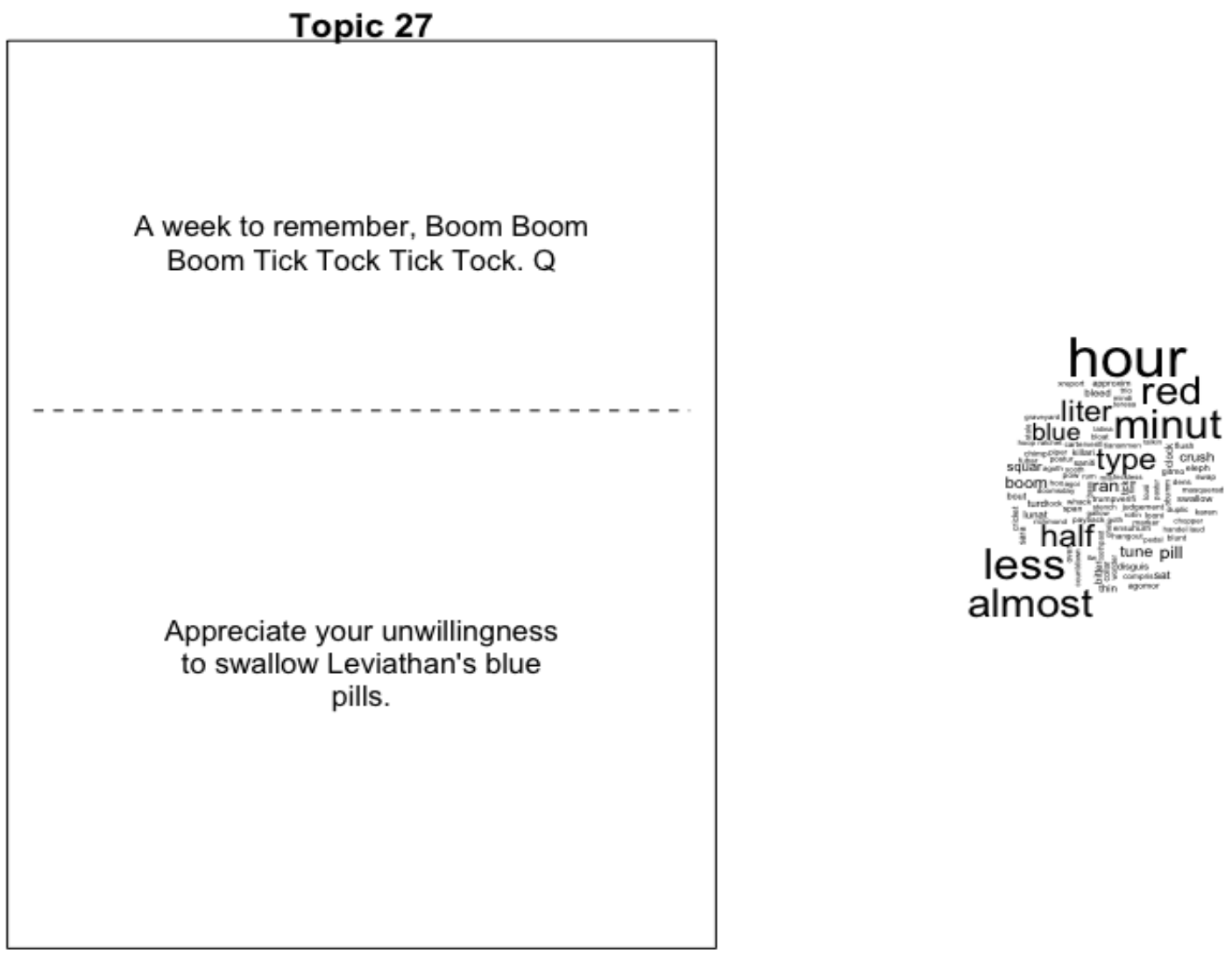

Figure 12: Visualization of nodal topic 27

Similarly, topic 27 concerns the popular alt-right terms "blue pill" and "red pill." The two metaphors come from the movie *The Matrix* (1999), where the hero (Neo) is offered two pills: a red pill, and a blue pill. The former refers to truth and knowledge, while the latter refers to ignorance. Neo chooses the red pill, and therefore, exposed to truth, leaves behind his deceived and ignorant life that is designed and controlled by the system. As previous studies have noted (Berger 2018; Lewis 2018; Stern 2019), to take the red pill implies a process in which a person is awakened to the alt-right-provided truth. Gab users utilize this iconic moment from a mainstream movie to signify the ways in which the alt-right movement and ideologies exposed them to the truth. In this case, this truth often supports larger narratives of racism, sexism, and other forms of oppression (Lewis 2018).

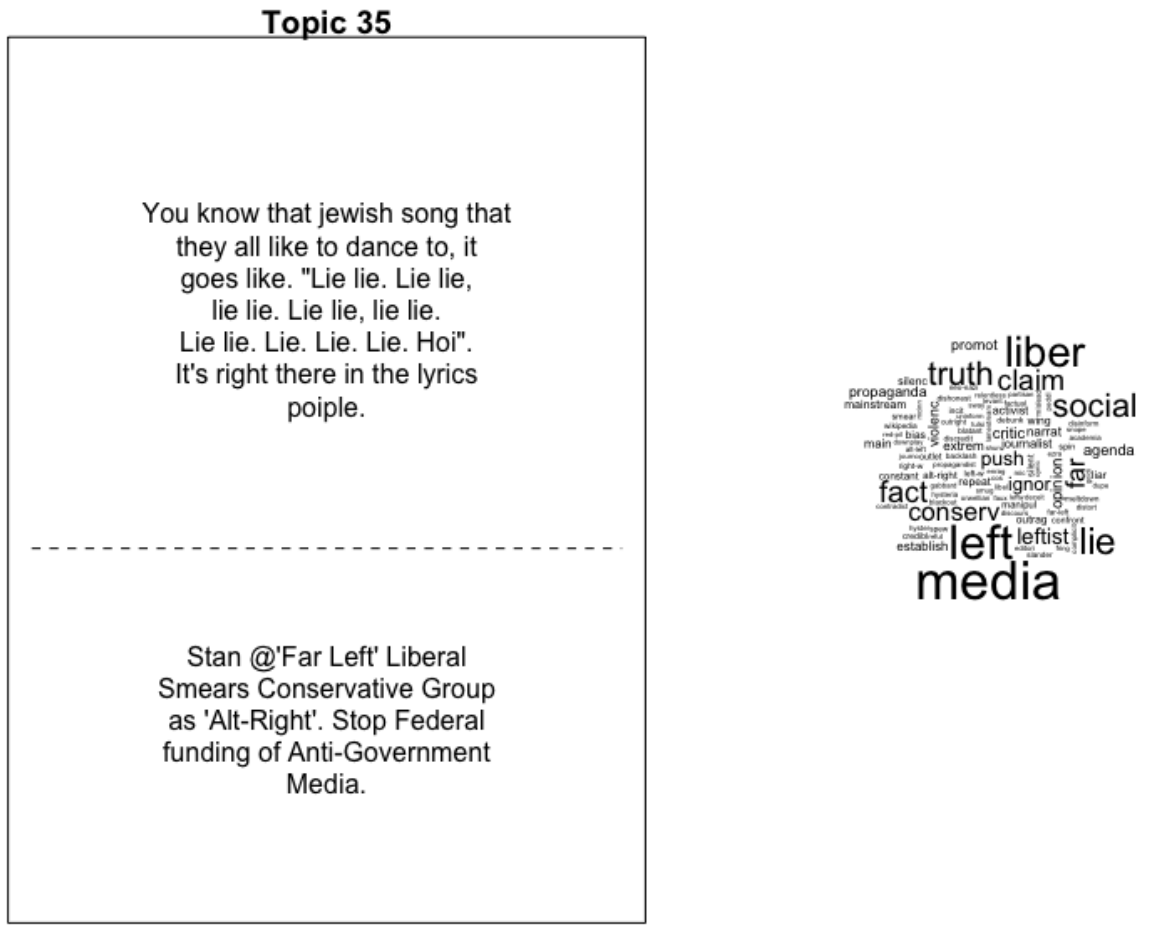

Figure 13: Visualization of nodal topic 35

Lastly, topic 35 is the general denigration of traditional media sites and purveyors as "mainstream media," a derogatory term suggesting that these outlets refuse to present views like the alt-right's as a way to silence their voices. This topic is perhaps the most salient from a content moderation perspective: it indicates that attempts to change the type of information presented to Gab users (for example, as way to address topics 19 and 27) would be difficult if the information hails from sources not trusted by the community. In other words, topic 35 raises the question "How do we challenge their truths?"

## CONCLUSION

The distinction between alt-right content and what alt-right users generate on Gab is blurred. Our results show that Gab, as a whole, is an alt-right platform, and Gab is aligned with the alt-right ideology. Similar to other social scientists' findings (Stern, 2019), the dominant discourses on Gab resemble selective alt-right forums such as sub-reddit TheRedPill. Thus, one could conclude that Gab is an alt-right platform, rather than simply a right-leaning platform that attracts a large number of alt-right users.

Previous research on Gab (Lima et al, 2018; McIlroy-Young & Anderson, 2019) has only categorized Gab as a right-leaning platform. In contrast, we provide a finer categorization of Gab as an alt-right platform. Our findings are crucial in the context of the increasing political and social discussion on the alt-right movement. Our results are based on topic models which provide in-depth topic analysis compared to prior work which used simple methods such as toxicity score and trigram. Specifically, our approach can generate meaningful topics like the "red-pill" that are absent in prior work. Topics such as these are crucial in classifying Gab as an alt-right platform. Our contribution also includes a detailed discussion of the generated topics along with their contextualized information (i.e., actual posts).

In this study, we examined popular topics of interest surrounding the platform's greater freedom of speech and minimal-to-no content moderation. These features are attractive to these users and have become topics of interest themselves, as users discussed the possibilities of having greater access to free speech. Such a perspective may have also contributed to the other observed topics, such as Holocaust denial, which may not be as explicit within the mainstream social media ecosystem, of which they distrust. Gab's gain in popularity during the 2016 U.S. presidential campaign (Lima et al. 2018) is echoed in enduring support and interest in Trump's presidency.

The use of the hashtag #Gabfam indicates that a form of community building is happening within Gab. This contrasts with past assumptions that white extremist events are lone-wolf attacks (Shapiro 2019; Berger 2019). Understanding how a community is formed and maintained within this platform seems to be crucial to implementing interventions. Given the ways in which these communities exist within an echo-chamber of alt-right ideologies that do not respond well to mainstream and alternate news sources, understanding how they build upon each other may help us understand the process of red-pilling. In addition, the function of bots on Gab, especially ones that might be replicating the most discussed topics may add to this process of community building and echo-chamber circulation.

The example of *The Matrix* and the process of red-pilling also speaks to the ways in which the alt-right has been generating cultural content that engages with main-stream culture. The remixing of and engagement with mainstream cultural products in addition to a production of their own speaks to the manner in which the alt-right interacts with the mainstream (Heikkilä's 2017). These products serve as a way to engage in meta-politics. They are more interested in receiving attention than with the politics of their ideologies.

Future work on Gab users and content should consider the ways in which disinformation is generated and shared. Reflecting the topic of Holocaust denial, it would seem that within an echo-chamber the function of disinformation is intensified. Considering Gab users' general distrust for mainstream media, future work on how Gab intersects with disinformation will be important. Because of the intellectualism of Gab users, it is inadequate to dismiss their rhetoric and ideological concerns. Dismissing the alt-right as unintelligent attention-seekers is detrimental to the work of intervention and disruption. As cautioned by Gökarıksel et al. (2019), we must work with our discomfort to understand how these ideas manifest. We need to consider the narratives produced in the intersections of disinformation, intellectualism, and the minimal-to-no content moderation found on Gab. Ignoring it is no longer an option, especially because of the violent real-life consequences that can emerge from such spaces.

# APPENDIX A

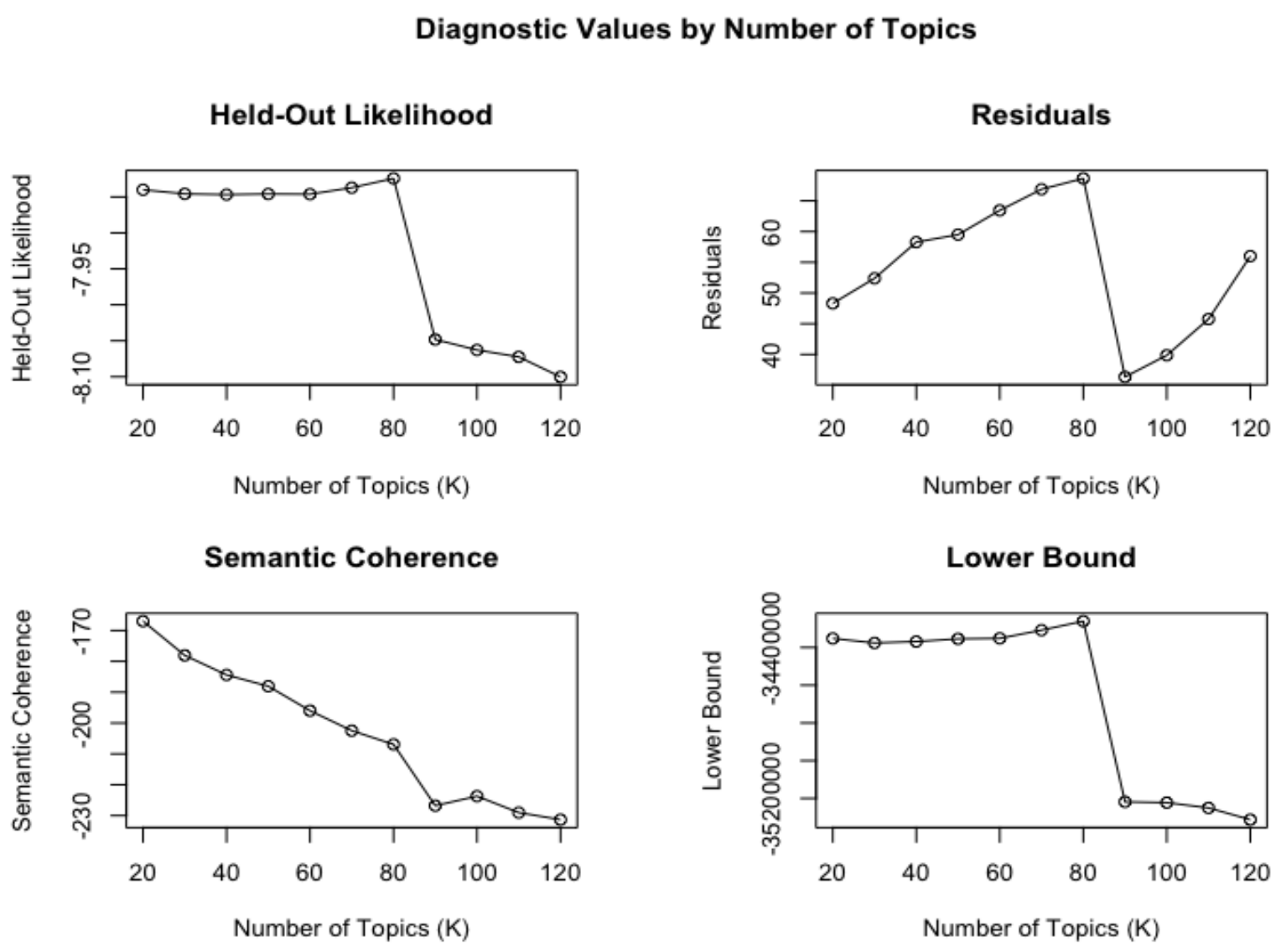

Figure A1: Diagnostics for selection of k number of topics

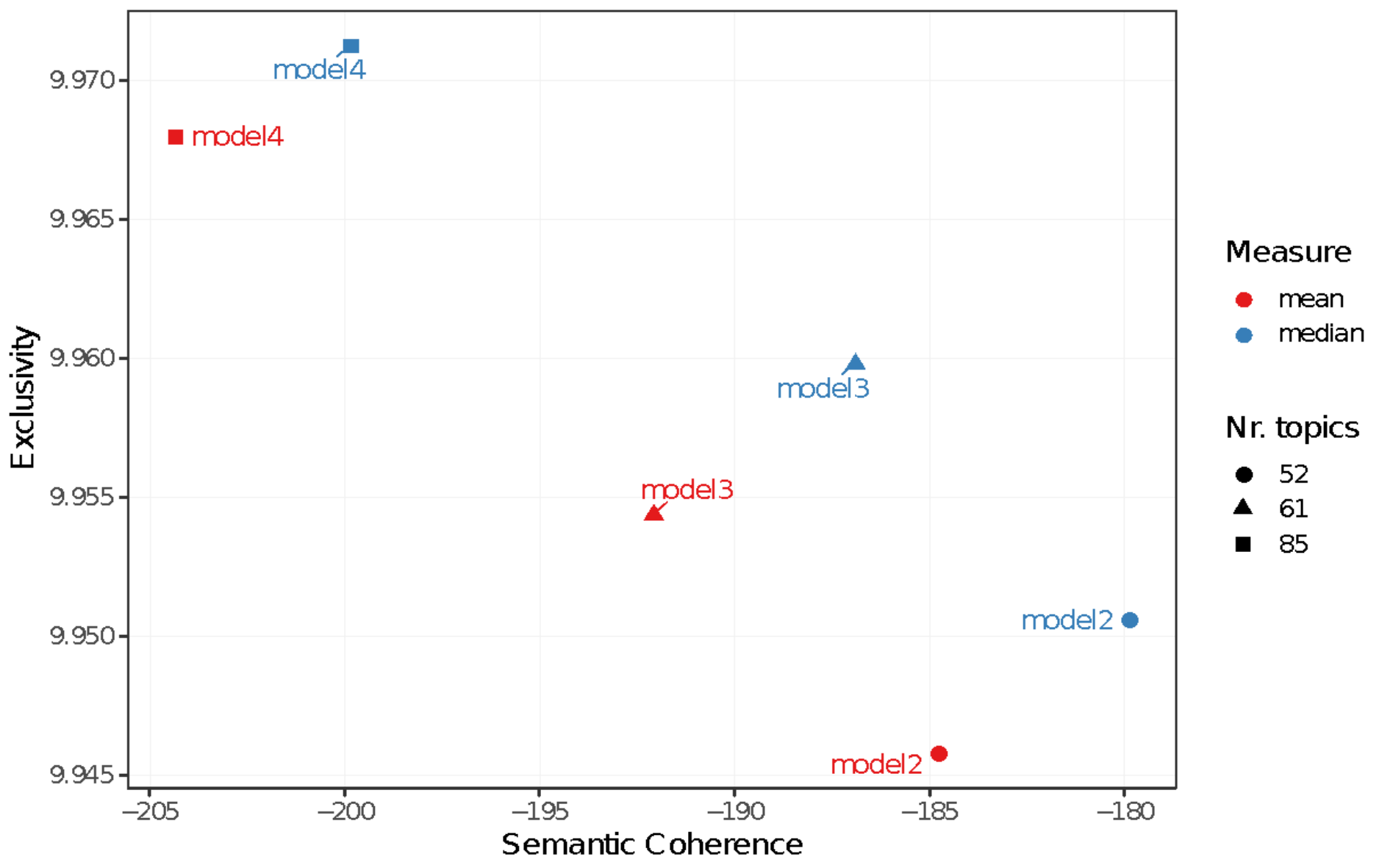

Figure A2: Final model comparison of GAB STM